# A method for virtual optical sectioning and tomography utilizing shallow depth of field

TIMUR E. GUREYEV,[1,2,*] HARRY M. QUINEY,[1] AND LESLIE J. ALLEN[1]

[1]*School of Physics, the University of Melbourne, Parkville, VIC 3010, Australia*
[2]*School of Physics and Astronomy, Monash University, Clayton, VIC 3800, Australia*
*\*timur.gureyev@unimelb.edu.au*

**Abstract:** A method is proposed for high-resolution, three-dimensional reconstruction of internal structure of objects from planar transmission images. The described approach can be used with any form of radiation or matter waves, in principle, provided that the depth of field is smaller than the thickness of the sample. The physical optics basis for the method is elucidated and the reconstruction algorithm is presented in detail. A simulated example demonstrates an application of the method to three-dimensional electron transmission imaging of a nanoparticle under realistic radiation dose and spatial resolution constraints. It is envisaged that the method can be applicable in high-resolution transmission electron microscopy, soft X-ray microscopy, ultrasound imaging and other areas.

## 1. Introduction

Multiple optical methods have been developed for three-dimensional (3D) imaging of internal structure of objects by means of computer processing of two-dimensional (2D) transmission images obtained with electrons, X-rays, visible light, ultrasound, etc. [1-3]. Computed tomography (CT) [4] is probably the best known example of a general method for such computational imaging. In CT, the contrast in the registered 2D images (projections) is assumed to be the result of variable absorption or phase shifts of the incident radiation as it propagates through the sample along straight rays through different parts of the sample. This interpretation of the images relies on the assumption that the changes in the propagation direction as a result of scattering inside the sample are sufficiently small and can be safely neglected. The latter condition is equivalent to the curvature of the Ewald sphere being negligible in the imaging setup, allowing one to replace the relevant part of the sphere with a tangential plane in the imaging model [2]. The flatness of the Ewald sphere is in turn equivalent to the depth of field (DOF) being larger than the thickness of the sample [1,5,6]. Slightly different definitions of the DOF can be found in the literature, but the one that is most relevant to our context is $DOF \cong \Delta^2/(2\lambda)$ [7], where $\Delta$ is the spatial resolution and $\lambda$ is the radiation wavelength (see the next section for details). The DOF may become smaller than the thickness of a typical sample e.g. in high-resolution transmission electron microscopy (TEM), soft X-ray microscopy or ultrasound imaging. In a high-resolution electron cryo-microscopy experiment one may have $\lambda \cong 0.02 \text{Å}$ (for 300 keV electrons) and $\Delta \cong 1 \text{Å}$, meaning that $DOF \cong 25 \text{Å}$, which is smaller than the size of many biological molecules of interest [8]. In soft X-ray microscopy, working in the so-called "water window", one may have $\lambda \cong 2.5 \text{nm}$ and $\Delta \cong 30 \text{nm}$, giving $DOF \cong 360 \text{nm}$, which is again smaller than the size of many relevant samples [9]. The usual experimental strategy in such cases is to try to increase the DOF so that the conventional straight-ray imaging model would be applicable and the corresponding CT-type methods could be used for the 3D reconstruction. However, in this paper we consider the opposite approach which explicitly uses the shallow DOF as a means for improving the spatial resolution and signal-to-noise (SNR) in the 3D reconstruction by a technique based on Diffraction Tomography (DT) [10-12]. This approach is similar in spirit to confocal microscopy [6,13] which utilises the shallow DOF for optical sectioning by essentially limiting the information in a given image to a thin transverse slice of the sample. However, unlike confocal microscopy and other techniques implementing some form of optical sectioning in hardware, the computational imaging technique considered in the present paper allows for a virtual optical sectioning (VOS) in software, while utilizing as input conventional 2D transmission images obtained without any special optical elements. We show that by applying this type of VOS it is sometimes possible to perform 3D imaging of suitable samples from a single 2D transmission image. More generally, these techniques may allow one to significantly reduce the sampling requirements of conventional CT in regards to the number of different projections (views) required for unambiguous 3D reconstruction with a given spatial resolution and SNR. The physical reasons behind these advantages over conventional CT approaches are rather straightforward and will be discussed next. The discussion serves as a justification for and an introduction to the "variable CTF" (vCTF) method that is developed in detail in the next section of this paper. However, many of the points discussed below are relevant to any DT-type technique, including, for example, Differential Holographic Tomography (DHT) [14] or Conjugated Holographic Reconstruction (CHR) [15].

Consider the fact that in the course of a conventional CT reconstruction the 2D contrast distribution in each planar image (projection) is effectively uniformly spread (numerically back-projected) over the whole reconstructed 3D volume along the straight lines extending from each detector pixel parallel to the illumination direction. This means that a "partial reconstruction" from a single 2D image, while having a non-trivial transverse spatial resolution as determined by the characteristics of the imaging system, at the same time has no longitudinal spatial resolution at all. An approximately isotropic 3D spatial resolution appears in CT reconstruction only after addition of sufficiently many back-projected partial reconstructions corresponding to different illumination directions (views). Let us call a one-dimensional (1D) back-propagated trace obtained in the process of 3D reconstruction, from the image contrast value in a given detector pixel of an input 2D image, a longitudinal point-spread function (LPSF), $L_{x,y}(z)$, where $\mathbf{r} = (x, y, z)$ are the Cartesian coordinates in 3D space, the transverse coordinates $(x, y)$ denote the position of the detector pixel and the longitudinal coordinate $z$ corresponds to the illumination direction (optic axis). Note that the term "point-spread function" becomes literal in this context. It is important to appreciate that this point-spread function (PSF) refers to the reconstruction operation, rather than to the initial (forward) imaging process. Accordingly, in the case of CT, the LPSFs will look like straight lines having the same value at any point $z$ along the optic axis. The CT LPSFs have such a form because, when the DOF is much larger than the sample thickness, an individual 2D image contains no information about the variation of the sample properties along the optic axis. Indeed, consider that in CT, the image contrast at each pixel corresponds to a line integral of the sample's refractive index, according to Beer's law [4]. Therefore, the sample's refractive index is effectively integrated along each such line into a single value registered in one pixel of the 2D image. Contrary to this, when the DOF is smaller than the sample thickness, a single transmission 2D image may already contain information about the variation of the refractive index along the illumination direction. This happens because the angular divergence of scattered rays can now be sufficiently large to imprint the information originating from different scattering centres located along the same illuminating ray into different pixels of the recorded 2D image. The information about the location of a scattering centre along the illumination direction, for example, is encoded simply by means of the variable "magnification" of diffraction patterns being proportional to the distance between the scattering centres and the image plane. In terms often used in electron and X-ray imaging, the contrast transfer function (CTF) here changes significantly over propagation distances comparable with the sample thickness [2,16]. Accordingly, in a reconstruction method properly designed for such imaging conditions, the back-propagated LPSFs will have non-trivial variation along the optic axis within the reconstructed sample volume. More specifically, in the vCTF method developed later in this paper, as in other DT methods in general, the reconstruction is performed according to the Fresnel back-propagation from the image plane(s) into the reconstruction volume containing the imaged sample. This process reverses the "forward" free-space propagation that took place in the process of image formation. It is then not unreasonable to expect that the vCTF LPSFs may have peaks at the locations of strong scattering centres, such as individual atoms in the case of high-resolution TEM [15]. The following simple example confirms such a conjecture.

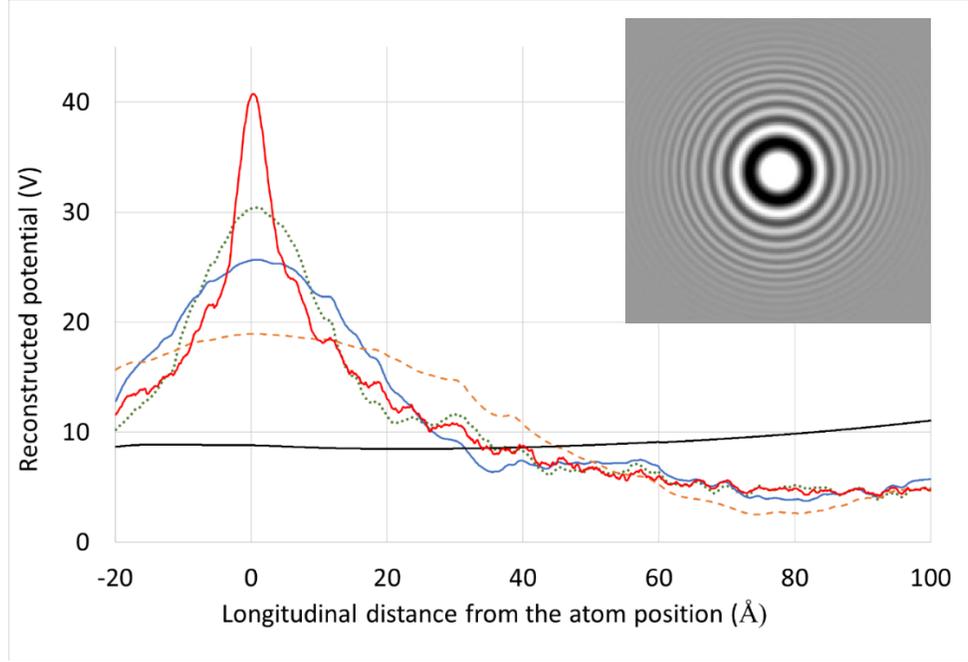

Fig. 1. LPSFs $L_{0,0}(z)$ of the vCTF reconstruction in the case of TEM imaging of a single carbon atom located at $z = 0$ and illuminated by a plane electron wave with the energy E = 300 keV. The image was collected at the defocus distance $z = 200$ Å at different effective spatial resolutions equal to $\Delta = 2$ Å (solid black curve), 1 Å (dashed orange curve), 0.66 Å (solid blue curve), 0.5 Å (dotted green curve) and 0.2 Å (solid red curve). The insert shows the defocused image with 0.5 Å spatial resolution. The corresponding DOFs are equal to approximately 102 Å, 25 Å, 11 Å, 6 Å and 1 Å, respectively. This figure demonstrates that, while the peaks of the LPSFs at the location of the scattering centre become sharper at higher spatial resolutions, the location of the maximum is independent of the resolution.

Consider a single carbon atom located at the origin of Cartesian coordinates in 3D space, $(x,y,z) = (0,0,0)$. The atom is illuminated by a plane monochromatic electron wave with the energy E = 300 keV propagating along $z$. The defocused 2D images in the plane $z = 200$ Å were calculated at different effective spatial resolutions using the publicly available software [17] which was partially based on the well-known "temsim" code [18-20]. Thermal vibrations with 0.1 Å RMS displacement at 300 K, adjusted to the temperature of 77 K, were incorporated into the calculations via a Debye-Waller factor. The simulated defocused images were then fed into the vCTF reconstruction algorithm (which is described in detail in the next section) and the corresponding back-propagated LPSFs at the central point of the images, i.e. $L_{0,0}(z)$, are shown in Figure 1. One can see that at the relatively "low" spatial resolution of 2 Å (i.e. at a relatively large DOF) the LPSF is almost flat and thus resembles the conventional back-projected LPSF of CT. However, at finer spatial resolutions the LPSFs develop a progressively sharper peak at the atom location, indicating the emergence of a non-trivial longitudinal spatial resolution which allows one to locate the scattering centre (the atom) in 3D from a single defocused image. It is also easy to see that the width of the peak is approximately equal to the DOF, which determines the longitudinal spatial resolution. The emergence of the peaks of the reconstructed signal at the location of strong scatterers is also improving the SNR of the reconstruction. Note that lower energies are often used in TEM in the context of materials sciences applications, e.g. E = 80 keV is a standard option on many aberration corrected EM systems these days. This will result in an approximately twice smaller DOF compared to 300 keV electrons, further improving the spatial resolution and the SNR of the 3D reconstruction using this approach.

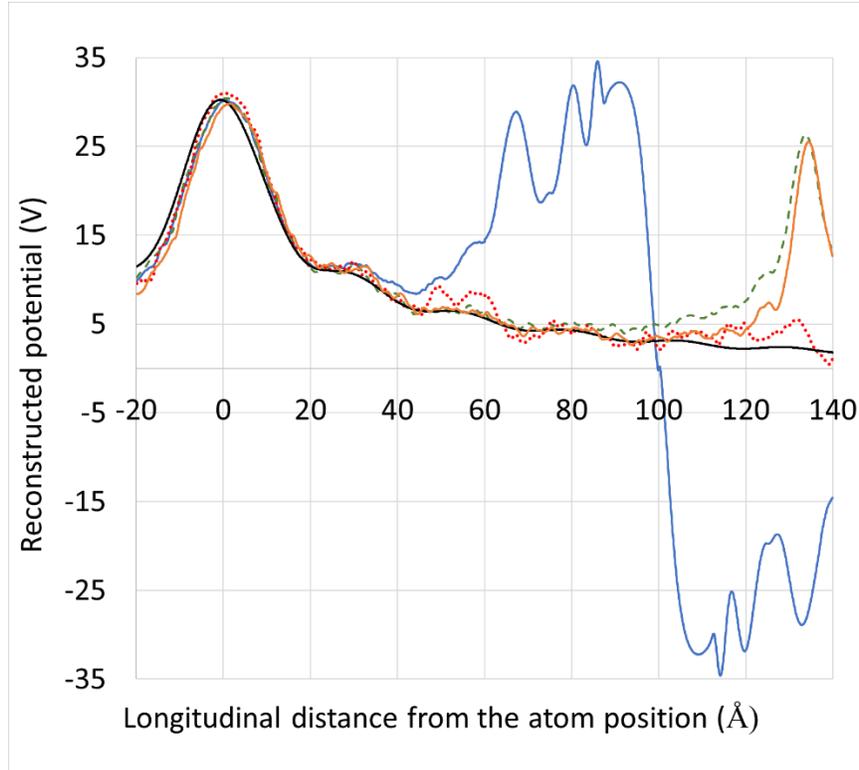

Fig. 2. LPSFs $L_{0,0}(z)$ of the vCTF reconstruction in the case of TEM imaging of a single carbon atom located at $z = 0$ and illuminated by a plane electron wave. The defocused images were collected at defocus distances equal to $z = 100$ Å (solid blue curve), 200 Å (dashed green curve), 200 Å with phenomenological absorption included (solid orange curve), and 5,000 Å (dotted red curve). Also plotted is the profile of the phase shift, $\varphi(0,0,z)$, corresponding to the scattering of the plane wave on the atomic potential (black curve) - see Section 2 for details. This figure demonstrates that a peak of the LPSFs always occurs at the location of the scattering centre, regardless of the position of the image plane.

Crucially, in a physically consistent DT-based technique, the reconstructed $z$-positions of the peaks of the LPSFs are invariant with respect to the distance between the sample and the image plane (Fig.1). This can be understood in simple terms by considering that the relative magnification of the Fresnel diffraction image of a given scattering centre is always proportional to the distance between the scattering centre and the image plane. Accordingly, a proper reconstruction method can intrinsically deduce the propagation distance to a given scattering centre from its contrast pattern in the defocused image. When several scattering centres are located within the sample at different positions along the illumination direction, the corresponding back-propagated LPSF peaks occur at the correct $z$-positions corresponding to the location of each centre. This important feature of DT-based reconstruction techniques is illustrated in two different ways in Figures 2 and 3. Note that similar ideas have been previously successfully exploited in the Big Bang Tomography method [21-23], even though the Big Bang Tomography technique is quite different in its implementation from the vCTF method developed in the present paper.

Figure 2 contains further results involving the scattering of a plane monochromatic electron wave on a single carbon atom, as described above in conjunction with Figure 1. The objective aperture here was equal to 40 mrad, corresponding to the spatial resolution $\Delta \cong 0.5$ Å, according to the convention used in TEM [5]. In this case, however, we varied the position of the image (defocus) plane, between $z = 100$ Å and $z = 5,000$ Å. In the case of $z = 5,000$ Å, a spherical aberration $C_3 = 2.7$ mm was also introduced, as used in some cryo-EM experiments [24]. One can see that in all the cases the reconstructed vCTF LPSF $L_{0,0}(z)$ had a similarly-shaped peak in the vicinity of the atom position, with the magnitude of $L_{0,0}(z)$ strongly decreasing away from the atom position until approximately half of the defocus distance. This indicates that, provided that the image plane is removed sufficiently far from the sample, a unique peak around an atomic position can be obtained in the reconstruction regardless of the defocus distance. Consequently, it may be also possible to find the longitudinal positions of the scattering centres inside the sample from a single 2D image, provided that different scattering centres are not shading each other in the image (see the discussion of this point below). In other words, it may be possible to perform a VOS under the described conditions, something that is clearly impossible when using conventional CT. The following example clarifies this point further.

We simulated a defocused image of the aspartate molecule, $C_4 H_7 N O_4$ [25], under the imaging conditions with a shallow DOF as described above, with objective aperture of 40 mrad and $C_3 = 0$. For the simulations, we centred the aspartate molecule within a $30 \times 30 \times 30$ Å$^3$ cube $Q_{30}$ located in the positive octant of the Cartesian coordinates $(x, y, z)$, with one corner at the point $(0, 0, 0)$ and all sides parallel to the coordinate axes. The simulation cube $Q_{30}$ was assumed to be illuminated by a plane monochromatic incident electron plane wave with an energy of 300 keV and uniform intensity equal to one everywhere, propagating along the $z$ axis. The multislice-based software code [17] was applied for the calculation of propagation of the electron wave through the sample, followed by the free-space propagation to the image plane. An image at $z = 17$ Å in Fig.3(a) shows that the size of the diffraction patterns of atoms is proportional to their distance from the image plane. The vCTF reconstruction was performed from a single defocused image simulated at $z = 115$ Å (Fig.3(b)). Cross-section through the 3D distribution of back-propagated LPSFs at two different planes, at $z = 10$ Å and $z = 22$ Å, are shown in Figs.3(c) and (d), respectively. The fact that different atomic positions are visible in the two planes confirms that VOS can be performed in this way. Analysing all back-propagated slices at different $z$ positions, the locations of all nine non-hydrogen atoms can be clearly identified in this reconstruction obtained from a single defocused image. The LPSF peaks corresponding to the individual atoms are elongated along the illumination direction. Nevertheless, by locating the highest points of the reconstructed peaks, the positions of all nine non-hydrogen atoms in the imaged molecule were retrieved with an average accuracy of 0.21 A and a maximum error of 0.46 Å. This confirms the hypothesis described above, i.e. that it may be possible to perform the VOS and quantitative 3D reconstruction of suitably sparse objects from a single transmission image, provided that the image was obtained with a DOF smaller than the thickness of the sample.

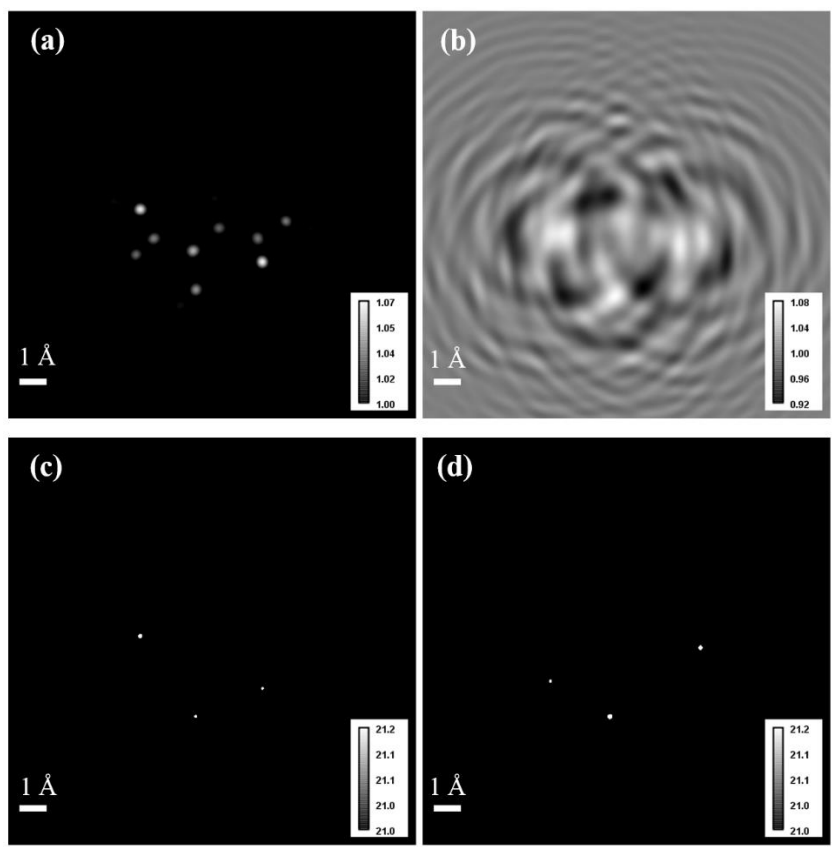

Fig. 3. (a) An image at $z = 17$ Å of the aspartate molecule obtained under the conditions of plane electron wave illumination. (b) An image at $z = 115$ Å used for the subsequent vCTF reconstruction. (c) A slice through the reconstructed 3D distribution at $z = 10$ Å. (d) A slice through the reconstructed 3D distribution at $z = 22$ Å. The grey-scale levels in (c) and (d) are normalized as the electrostatic potential (in volts) - see Section 2 for details.

As may already be evident from the previous example, a sufficient "sparsity" of the imaged object is a necessary condition for unambiguous 3D reconstruction from a single 2D image. When two or more scattering centres appear along the same line parallel to the illumination direction, one of them may effectively "shade" the other in the image and consequently, the VOS approach described above may fail. This issue corresponds to the known problem of multiple scattering being incompatible with the first Born and first Rytov approximations on which the DT approaches are typically based [8,10,14,26,27]. A reconstruction using DT-based methods may still be successfully performed in the presence of strong multiple scattering, as the example shown in section 3 below demonstrates. However, in order to achieve a good-quality 3D reconstruction in such cases, several transmission images need to be collected at different illumination directions and used in the reconstruction. When partial reconstructions obtained at each illumination direction are added together, the multiple scattering effects, which are usually strongly directional, tend to average out into a slowly-varying background. On the other hand, the single scattering signals, which tend to be more isotropic, usually add up constructively and produce enhanced signals at the location of the scattering centres. In this context, the mechanism of the vCTF and related methods, such as DHT or CHR, becomes closer to that of conventional CT. However, the shallow DOF can still provide substantial benefits in this case. Consider first that in order to obtain a high-quality CT reconstruction from images collected at a significant sample-to-detector distance it is usually necessary to compensate for the effects of free-space propagation between the sample and the detector on the image contrast. Note that a substantial sample-to-detector distance is typically required in the first instance in order to achieve a sufficiently strong image contrast when imaging weakly absorbing or pure phase objects [3]. The "compensation" for the free-space propagation during the 3D reconstruction is typically performed by means of various phase retrieval methods [3,28,29] which allow one to reconstruct the complex amplitude in the image plane and subsequently carry out a numerical free-space back-propagation along the optic axis to the position of the imaged object. In TEM imaging, such methods are usually referred to as "CTF correction", which is in essence a phase retrieval procedure applicable to weak phase objects [2,8,16]. A problem occurs with these methods in the case of shallow DOF due to the fact that the back-propagation (CTF correction) is performed at each illumination direction with respect to a single plane in the reconstruction volume (usually, the central plane of the sample). When the DOF is smaller than the thickness of the sample, the Fresnel propagation within the sample becomes significant. Consequently, the back-propagation to the central plane of the sample becomes inaccurate for the scattering centres, such as atoms, that are located on the periphery of the sample at distances exceeding the DOF from the central plane. In effect, for such peripheral scattering centres, the back-propagation (CTF correction) is then performed using an incorrect defocus distance. A typical consequence of this error in the propagation distance is the resultant blurring of the reconstructed distribution of the refractive index in the vicinity of such peripheral centres. In other words, this leads to the loss of 3D spatial resolution in the reconstruction of the peripheral regions of the sample. The DT-based methods, such as vCTF, which intrinsically take the Fresnel propagation within the sample into account, do not suffer from the same loss of spatial resolution. The following simple example illustrates this point. A more detailed demonstration of the same effect can be found in section 3.

The final example of this section was calculated for a single carbon atom placed at two different positions: at the origin of Cartesian coordinates in 3D space, $(x, y, z_0) = (0,0,0)$, and at $(x, y, z_0) = (0,0,50\,\text{Å})$, i.e. at a distance of 50 Å from the origin. In each of the two cases, the atom was illuminated by a plane monochromatic electron wave under the same imaging conditions as described above and a defocused 2D image at $z = 200\,\text{Å}$ was calculated. The defocused images were then fed into the vCTF reconstruction algorithm and into the conventional CTF correction (cCTF) algorithm (see the details about these algorithms in the next section), both using the same defocus distance of $200\,\text{Å}$ as an input parameter. Figure 4 shows the transverse (along the *y* coordinate) cross-sections through the resultant reconstructed signals at the two different *z*-positions of the atom. One can see that while the vCTF and the cCTF results were identical for the atom located at $z = 0$, the cCTF result for the peripheral atom located at $z = 50\,\text{Å}$ showed significant blurring and depression, i.e. it exhibited a clear loss of spatial resolution. In contrast to this, the vCTF reconstructed signal for the atom at $z = 50\,\text{Å}$ had virtually the same profile as that for the centrally-located atom. The latter was made possible because the vCTF method intrinsically uses the contrast produced by a given atom in the defocused image to determine the "correction" of the defocus distance relative to the "average" defocus distance given as input. This outcome confirms the statement made above, i.e. that in the case of a shallow DOF the conventional CTF correction methods are likely to experience a loss of spatial resolution, while the methods based on the DT approach do not suffer from the same problem. The latter benefit may become important, for example, in atomic-resolution cryo-EM, where alternative methods for Ewald sphere correction are already being used [30,31]. It remains to be seen if the vCTF method can be advantageous, at least in some circumstances, compared to the previously published methods, in terms of its stability, accuracy or the ease of application. Applications in other imaging areas, using e.g. soft X-rays, may also potentially benefit from the use of this type of method.

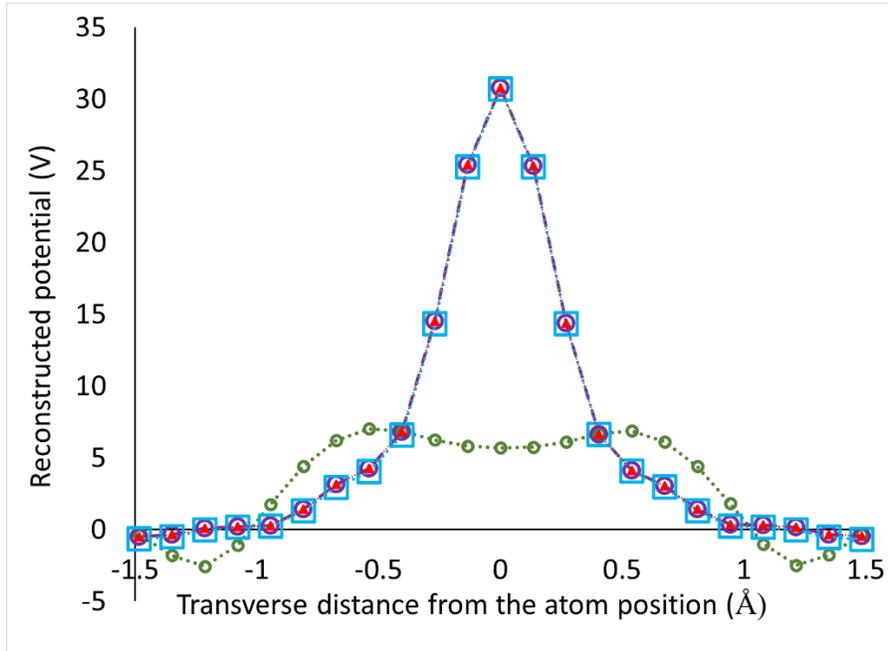

Fig. 4. Transverse cross-sections through the reconstructed vCTF signal in the case of TEM imaging of a single carbon atom located at $z_0 = 0$ or $z_0 = 50$ Å and illuminated by a plane electron wave. The reconstructions were performed from a single 2D image at $z = 200$ Å. Different lines show the results for the atom at $z_0 = 0$, using the vCTF method (red line with triangular markers) and the conventional CTF-correction method (purple line with circular markers), and for the atom at $z_0 = 50$ Å, using the vCTF method (blue line with square markers) and the conventional CTF-correction method (green line with circular markers). This figure demonstrates that, in the case of a shallow DOF, the conventional CTF-corrected reconstruction may experience significant blurring and depression of the reconstructed signal near peripherally-located scattering centres, compared to the centrally located centres, while the vCTF method does not suffer from the same problem.

The structure of the rest of this paper is as follows. The next section presents a theoretical description of the proposed vCTF method for VOS and CT reconstruction of weakly scattering objects. A simulated example of application of this method to 3D imaging of a nanoparticle is given in section 3. Section 4 contains a final brief discussion and conclusions.

## 2. Theory of variable-distance CTF correction (vCTF) method

Consider an imaging scheme where a monochromatic plane wave $U_{in} \equiv I_{in}^{1/2} \exp(i2\pi z / \lambda)$ with a uniform intensity $I_{in} = \text{const}$ illuminates a weakly scattering object. The complex amplitude $U(\mathbf{r})$, $\mathbf{r} = (\mathbf{r}_\perp, z)$ of the wave inside the object satisfies the stationary wave equation $\nabla^2 U(\mathbf{r}) + [2\pi n(\mathbf{r}) / \lambda]^2 U(\mathbf{r}) = 0$, where $n(\mathbf{r}) = 1 + \delta(\mathbf{r}) + i\beta(\mathbf{r})$ is the refractive index and the dependence of all functions on $\lambda$ has been made implicit for brevity. The assumed weakness of the scattering object means that $|\delta(\mathbf{r})| \ll 1$ and $|\beta(\mathbf{r})| \ll 1$. In the case of electron microscopy, one has $n(\mathbf{r}) \cong 1 + V(\mathbf{r}) / (2E)$, where $V(\mathbf{r}) \geq 0$ is the electrostatic potential, $E$ is the accelerating voltage and $V(\mathbf{r}) / (2E)$ is typically much less than 1 (see e.g. [18]). We consider the problem of reconstruction of the 3D distribution of $n(\mathbf{r}) - 1$ from the intensity of transmitted waves measured at some distance(s) from the object.

Let us assume that the imaged sample can be represented as a collection of highly localized "scattering centres" which can be treated independently in the paraxial first Born approximation. In other words, we assume that each individual scattering centre is illuminated by an unperturbed incident plane wave and that each scattered wave travels towards the detector without further interaction with the sample. Accordingly, we represent the 3D distribution of the refractive index in the sample as a sum of functions localized in the vicinity of individual scattering centres:

$$n(\mathbf{r}) - 1 = \sum_{m=1}^{M} \delta n_m (\mathbf{r} - \mathbf{r}_m) . \qquad (1)$$

Here $\mathbf{r}_m = (\mathbf{r}_{m,\perp}, z_m)$ are the locations of the scattering centres and the functions $|\delta n_m(\mathbf{r})| \ll 1$ are localized around $\mathbf{r} = 0$. In particular, such a model is suitable for atomistic representation of molecules or nanoparticles, but it can have much more generic applicability, in principle. For each individual scattering centre, the local effect on the incident wave can be expressed in terms of a linear integral of the local increment of the refractive index, $\delta n_m (\mathbf{r} - \mathbf{r}_m)$. As a result, the interaction of the incident wave with the whole sample contained inside the slab $-A \leq z \leq A$ can be represented via the first Born approximation applied to the transmission function of each scattering centre: $T(\mathbf{r}_\perp, z_m) = \exp[i\Phi_m(\mathbf{r}_\perp)] \cong 1 + i\Phi_m(\mathbf{r}_\perp)$, with

$$\Phi_m(\mathbf{r}_\perp) = (2\pi/\lambda) \int_{-A}^{A} \delta n_m(\mathbf{r}_\perp - \mathbf{r}_{m,\perp}, z - z_m) dz . \qquad (2)$$

After the interaction of the plane incident wave with the sample, the whole complex amplitude in the half-space $z > A$, consisting of the incident wave and the singly-scattered waves, can be expressed via the usual Fresnel diffraction integrals:

$$U(\mathbf{r}_\perp, z) \cong U_{in}(\mathbf{r}_\perp, z)\{1 + \sum_{m=1}^{M} F(\mathbf{r}_\perp, z - z_m) * (i\Phi_m)(\mathbf{r}_\perp)\}, \qquad (3)$$

where $F(\mathbf{r}_\perp, z) = [1/(i\lambda z)] \exp[i\pi r_\perp^2/(\lambda z)]$ is the Fresnel propagator and the asterisk denotes the 2D convolution with respect to transverse coordinates. In accordance with eq.(2), $\Phi_m(\mathbf{r}_\perp) = \varphi_m(\mathbf{r}_\perp) + i a_m(\mathbf{r}_\perp)$, where $\varphi_m(\mathbf{r}_\perp) \equiv (2\pi/\lambda) \int_{-A}^{A} \delta_m(\mathbf{r}_\perp - \mathbf{r}_{m,\perp}, z - z_m) dz$, $a_m(\mathbf{r}_\perp) \equiv (2\pi/\lambda) \int_{-A}^{A} \beta_m(\mathbf{r}_\perp - \mathbf{r}_{m,\perp}, z - z_m) dz$, with both $\varphi_m(\mathbf{r}_\perp)$ and $a_m(\mathbf{r}_\perp)$ being much smaller than unity. Expressing the complex amplitude $U(\mathbf{r}_\perp, z)$ on the left-hand side of eq.(3) as $U(\mathbf{r}_\perp, z) = U_{in}(\mathbf{r}_\perp, z) \exp[-a(\mathbf{r}_\perp, z) + i\varphi(\mathbf{r}_\perp, z)] \cong U_{in}(\mathbf{r}_\perp, z)[1 - a(\mathbf{r}_\perp, z) + i\varphi(\mathbf{r}_\perp, z)]$, we can write the following system of linear equations for the real and imaginary parts of eq.(3):

$$\begin{cases} a(\mathbf{r}_\perp, z) = \sum_{m=1}^{M} \left( \dfrac{\sin\{\pi r_\perp^2/[\lambda(z - z_m)]\}}{\lambda(z - z_m)} * a_m(\mathbf{r}_\perp) - \dfrac{\cos\{\pi r_\perp^2/[\lambda(z - z_m)]\}}{\lambda(z - z_m)} * \varphi_m(\mathbf{r}_\perp) \right) \\ \varphi(\mathbf{r}_\perp, z) = \sum_{m=1}^{M} \left( \dfrac{\cos\{\pi r_\perp^2/[\lambda(z - z_m)]\}}{\lambda(z - z_m)} * a_m(\mathbf{r}_\perp) + \dfrac{\sin\{\pi r_\perp^2/[\lambda(z - z_m)]\}}{\lambda(z - z_m)} * \varphi_m(\mathbf{r}_\perp) \right) \end{cases} . \quad (4)$$

Taking the 2D Fourier transform, $\hat{f}(\mathbf{q}_\perp) \equiv \iint \exp(-i 2\pi \mathbf{q}_\perp \mathbf{r}_\perp) f(\mathbf{r}_\perp) d\mathbf{r}_\perp$, of eqs.(4), we can also obtain

$$\begin{cases} \hat{a}(\mathbf{q}_\perp, z) = \sum_{m=1}^{M} \{\cos[\pi\lambda(z-z_m)q_\perp^2]\hat{a}_m(\mathbf{q}_\perp) - \sin[\pi\lambda(z-z_m)q_\perp^2]\hat{\varphi}_m(\mathbf{q}_\perp)\} \\ \hat{\varphi}(\mathbf{q}_\perp, z) = \sum_{m=1}^{M} \{\sin[\pi\lambda(z-z_m)q_\perp^2]\hat{a}_m(\mathbf{q}_\perp) + \cos[\pi\lambda(z-z_m)q_\perp^2]\hat{\varphi}_m(\mathbf{q}_\perp)\} \end{cases} \quad (5)$$

The latter equations correspond to the Weak Object Approximation (WOA) to the Fresnel diffraction integral [2]. Results similar to eq.(5) with $M = 1$ can be found in earlier publications, e.g. [16]. These results have been previously applied for the recovery of object-plane phase and intensity from defocused images, see e.g. [29] and further references therein. However, in the present case we would like to make an extra step and recover the individual components $a_m(\mathbf{r}_\perp)$ and $\varphi_m(\mathbf{r}_\perp)$, rather than just the "bulk" terms $a(\mathbf{r}_\perp, 0) \cong \sum_{m=1}^{M} a_m(\mathbf{r}_\perp)$ and $\varphi(\mathbf{r}_\perp, 0) \cong \sum_{m=1}^{M} \varphi_m(\mathbf{r}_\perp)$. The latter bulk terms emerge from eqs.(4) when all $z_m$ are set to zero, i.e. when the Fresnel propagation of the scattered waves inside the sample is ignored, or more precisely, when it is approximated by the propagation over a common "average" distance for all scatterers regardless of their actual positions inside the sample.

## 2.1. Back-propagated vCTF signal in the case of a weak pure-phase object

In order to simplify the following derivation, we shall concentrate first on the Weak Phase Object Approximation (WPOA) case, in which $a_m(\mathbf{r}_\perp) = 0$ for all $m$, i.e. the attenuation is negligibly small. We will present an extension of the results to a case involving non-trivial attenuation afterwards. We will also explain why the resultant reconstruction method is applicable to a broader class of samples that does not necessarily satisfy the WOA.

Let us consider first the case of a single weak phase scattering centre, so that the first of eqs.(5) becomes $\hat{a}(\mathbf{q}_\perp, z) = -\sin[\pi\lambda(z-z_m)q_\perp^2]\hat{\varphi}_m(\mathbf{q}_\perp)$, where $m$ is an arbitrary "place-holder" index. Note that the intensity is equal to $I(\mathbf{r}_\perp, z) = I_{in}\exp[-2a(\mathbf{r}_\perp, z)] \cong I_{in}[1 - 2a(\mathbf{r}_\perp, z)]$, and hence the contrast function, $K(\mathbf{r}_\perp, z) \equiv 1 - I(\mathbf{r}_\perp, z)/I_{in} = 2a(\mathbf{r}_\perp, z)$, satisfies the following equation:

$$\hat{K}(\mathbf{q}_\perp, z) = 2\sin[\pi\lambda(z_m - z)q_\perp^2]\hat{\varphi}_m(\mathbf{q}_\perp). \quad (6)$$

We assume the defocus plane to be $z = D$ and define a function to be used as a Fresnel back-propagated "signal" in the proposed vCTF reconstruction method in the case of weak phase objects:

$$\hat{f}(\mathbf{q}_\perp, z) \equiv \frac{\hat{K}(\mathbf{q}_\perp, D)}{2\sin[\pi\lambda(z-D)q_\perp^2]}. \quad (7)$$

The LPSFs of the vCTF reconstruction considered in section 1 above generally correspond to the inverse Fourier transform of the "signal" defined in eq.(7), i.e. $L_{x,y}(z) \equiv f(x, y, z)$. More precisely, as eq.(7) has a singularity at $z = D$ and other points where the denominator is equal to zero, it needs to be regularized, e.g. by introducing a function $\hat{f}_\varepsilon(\mathbf{q}_\perp, z) \cong \hat{f}(\mathbf{q}_\perp, z)$ with a small positive constant $\varepsilon$, $0 < \varepsilon \ll 1$:

$$\hat{f}_\varepsilon(\mathbf{q}_\perp, z) \equiv \frac{\sin[\pi\lambda(z-D)q_\perp^2]}{2\sin^2[\pi\lambda(z-D)q_\perp^2]+\varepsilon}\hat{K}(\mathbf{q}_\perp, D). \quad (8)$$

When $|\pi\lambda(D-z)q_\perp^2| < \varepsilon$ (e.g. near the defocus plane), we have $\hat{f}_\varepsilon(\mathbf{q}_\perp, z) \cong \hat{K}(\mathbf{q}_\perp, D)\varepsilon^{-1}\pi\lambda(z-D)q_\perp^2$, which is an odd function of the argument $(z-D)$, i.e. it is antisymmetric with respect to the position, $z = D$, of the defocus plane and is equal to zero at the exact defocus position, $\hat{f}_\varepsilon(\mathbf{q}_\perp, D) = 0$. Note, however, that when the defocus plane is sufficiently far away from the volume containing the imaged object, the behaviour of the function $\hat{f}_\varepsilon(\mathbf{q}_\perp, z)$ near the defocus plane does not affect the reconstruction of the object.

## 2.2. Physical meaning of the vCTF reconstructed signal

When $z = z_m$ in eq.(7), we have, $\hat{f}(\mathbf{q}_\perp, z_m) = \hat{\varphi}_m(\mathbf{q}_\perp)$, according to eq.(6) with $z = D$. This corresponds to the phase retrieval in a conventional CTF-corrected reconstruction with a fixed defocus distance. In contrast to that, the defocused distance $z$ in eq.(8) can be varied, e.g. within the thickness of a volume known to contain the sample. For that reason, we call the method associated with eq.(8) the variable CTF (vCTF) reconstruction. According to the structure of eq.(7), for any $z$, $\hat{f}(\mathbf{q}_\perp, z)$ is equal to the Fourier transform of a hypothetical phase distribution, $\tilde{\varphi}(\mathbf{r}_\perp, z)$, in the plane $z$ that produces the given contrast function $K(\mathbf{r}_\perp, D)$ upon free-space propagation of the complex amplitude $\tilde{U}(\mathbf{r}_\perp, z) = I_{in}^{1/2} \exp[i2\pi z/\lambda + i\tilde{\varphi}(\mathbf{r}_\perp, z)]$ to the plane $D$. However, such a description does not tell us much about the behaviour that can be expected of the vCTF signal defined by eq.(7) or eq.(8). In order to investigate this behaviour, let us substitute $\hat{K}(\mathbf{q}_\perp, D) = 2\hat{\varphi}_m(\mathbf{q}_\perp) \sin[\pi\lambda(z_m - D)q_\perp^2]$ into eq.(7) and apply the identity $\sin[\pi\lambda(z_m - D)q_\perp^2] = \sin[\pi\lambda(z - D)q_\perp^2]\cos[\pi\lambda(z_m - z)q_\perp^2] + \cos[\pi\lambda(z - D)q_\perp^2]\sin[\pi\lambda(z_m - z)q_\perp^2]$. We can then re-write eq.(7) as $\hat{f}(\mathbf{q}_\perp, z) = \{\cos[\pi\lambda(z_m - z)q_\perp^2] + \cot[\pi\lambda(z - D)q_\perp^2]\sin[\pi\lambda(z_m - z)q_\perp^2]\}\hat{\varphi}_m(\mathbf{q}_\perp)$. Applying here eqs.(5) with a single term and with $\hat{a}_m(\mathbf{q}_\perp) = 0$, and regularizing both sides as in eq.(8), we obtain

$$\hat{f}_\varepsilon(\mathbf{q}_\perp, z) \cong \hat{\varphi}_m(\mathbf{q}_\perp, z) + \frac{\cos[\pi\lambda(z - D)q_\perp^2]\sin[\pi\lambda(z - D)q_\perp^2]}{2\sin^2[\pi\lambda(z - D)q_\perp^2] + \varepsilon}\hat{K}_m(\mathbf{q}_\perp, z). \quad (9)$$

Here $\hat{K}_m(\mathbf{q}_\perp, z)$ and $\hat{\varphi}_m(\mathbf{q}_\perp, z)$ are the intensity contrast and the phase, respectively, which are observed upon the free-space propagation of the complex amplitude $I_{in}^{1/2} \exp[i2\pi z/\lambda + i\varphi_m(\mathbf{r}_\perp)]$ from the plane $z_m$, containing our weak phase scatterer with index $m$, to an arbitrary plane $z$. The second additive term on the right-hand side of eq.(9) is equal to zero, because, according to eq.(8) with the arguments $z$ and $D$ swapped, $\hat{K}_m(\mathbf{q}_\perp, z)\sin[\pi\lambda(z - D)q_\perp^2]/\{2\sin^2[\pi\lambda(z - D)q_\perp^2] + \varepsilon\} = -\hat{f}_\varepsilon(\mathbf{q}_\perp, D) = 0$. Thus, we arrive at the key equation

$$\hat{f}_\varepsilon(\mathbf{q}_\perp, z) \cong \hat{\varphi}_m(\mathbf{q}_\perp, z). \quad (10)$$

The free-space propagation of the phase and, hence, in accordance with eq.(10), the behaviour of the vCTF signal, is described by a simplified form of the second of eqs.(4):

$$f(\mathbf{r}_\perp, z) \cong \frac{\sin\{\pi r_\perp^2/[\lambda(z - z_m)]\}}{\lambda(z - z_m)} * \varphi_m(\mathbf{r}_\perp). \quad (11)$$

Equation (11) shows that the function $f(\mathbf{r}_\perp, z)$ is symmetrical in $z$ with respect to the point $z_m$ and has a maximum (peak) at $z = z_m$. As $|z - z_m|$ increases, $f(\mathbf{r}_\perp, z)$ can generally be expected to develop as an expanding "diffraction" pattern (oscillatory, as a function of $r_\perp$) with a decreasing magnitude. Since the right-hand side of eq.(11) does not depend on the position of the image plane $D$, the behaviour of the function $f(\mathbf{r}_\perp, z)$ in the vicinity of the scattering centre is independent of the distance *D* (as demonstrated earlier in Figs. 2 and 3), despite the fact that the vCTF signal $f(\mathbf{r}_\perp, z)$ is defined by eq.(8) with an explicit reference to the image plane.

Equation (11) also allows us to estimate the longitudinal spatial resolution of the vCTF method. When $r_\perp = \Delta/2$, where $\Delta$ represents the transverse spatial resolution as before (and hence the corresponding diameter is then $2r_\perp = \Delta$), the condition for the argument of the sine function in eq.(11) to be equal to $\pi/2$ [7] is $\Delta^2/(4\lambda z) = 1/2$, which corresponds to $z = \Delta^2/(2\lambda) = \text{DOF}$. Therefore, the longitudinal spatial resolution in the vCTF is approximately equal to the DOF.

We previously calculated analytically, using the first Born approximation, the free-space propagation of a plane electron wave scattered on a weak Gaussian electrostatic potential *V* localized at the origin of coordinates [15]. In that case, the phase function is also Gaussian, $\varphi_G(\mathbf{r}_\perp, 0) = [\pi/(\lambda E)] \int V(\mathbf{r}_\perp, z') dz' = \varepsilon G(\mathbf{r}_\perp)$, where $G(\mathbf{r}_\perp) \equiv \exp[-r_\perp^2/(2\sigma^2)]$ and $\varepsilon$ is a small constant, $0 < \varepsilon \ll 1$. It turns out that, for such a Gaussian phase function, the corresponding convolution in eq. (11) with $z_m = 0$ can be explicitly evaluated. The resultant complex amplitude propagating in free space has the phase equal to $2\pi z/\lambda + \varphi_G(\mathbf{r}_\perp, z)$, with the small "diffracting" Gaussian term

$$\varphi_G(\mathbf{r}_\perp, z) \cong \frac{\varepsilon \exp(-r_{\perp,z}^2)}{(1+b_z^2)} [\cos(b_z r_{\perp,z}^2) + b_z \sin(b_z r_{\perp,z}^2)], \qquad (12)$$

where $b_z = \lambda z/(2\pi\sigma^2)$ and $r_{\perp,z}^2 = r_\perp^2/[2\sigma^2(1+b_z^2)]$. Qualitatively, the behaviour of the function $\varphi_G(\mathbf{r}_\perp, z)$ is consistent with the general behaviour of the function $f(\mathbf{r}_\perp, z)$ as described above after eq.(11), with the maximum located near the position of the scatterer and the gradual diffraction spreading observed as the wave propagates along *z*.

Note also that eq.(11) transforms into the conventional CTF correction, when the convolution kernel in eq.(11) becomes close to the Dirac delta-function, $(\lambda z)^{-1} \sin[\pi r_\perp^2/(\lambda z)] \cong \delta(r_\perp)$. In the reciprocal space, this condition becomes $\cos(\pi \lambda z q_\perp^2) \cong 1$, or $\lambda T q_{\perp,\max}^2 \ll 1$, where *T* is the thickness of the sample and $q_{\perp,\max}$ is the largest detectable spatial frequency. This is precisely the condition for the Ewald sphere to be sufficiently flat. The same condition can be formulated in the real space in terms of the minimal Fresnel number, $N_{\min}^F \equiv \Delta^2/(\lambda T) \gg 1$, where $\Delta \cong 1/q_{\perp,\max}$ is the spatial resolution. The latter inequality represents a typical condition for the applicability of the geometrical optics [3], i.e. the straight rays approximation. Finally, this is also related to the condition for the applicability of CT methods as discussed in the Introduction, i.e. for the DOF to be larger than the sample thickness, $\text{DOF} = \Delta^2/(2\lambda) > T$, which is equivalent to $\lambda T q_{\perp,\max}^2 < 1/2$.

In the case of an object containing multiple weak phase scatterers, eq.(11) becomes

$$f(\mathbf{r}_\perp, z) \cong \sum_{m=1}^M \frac{\sin\{\pi r_\perp^2/[\lambda(z-z_m)]\}}{\lambda(z-z_m)} * \varphi_m(\mathbf{r}_\perp), \qquad (13)$$

where the vCTF signal $f(\mathbf{r}_\perp, z)$ is still defined by eq.(7), but the contrast function $\hat{K}(\mathbf{q}_\perp, D)$ now has contribution from all *M* scattering centres. When the phase functions $\varphi_m(\mathbf{r}_\perp)$ are highly localized, as assumed above, different terms of the sum in eq.(13) will have peaks (maxima) at different positions $z = z_m$, and in the vicinity of $\mathbf{r} = \mathbf{r}_m$ we will have $f(\mathbf{r}_\perp, z_m) \cong \varphi_m(\mathbf{r}_\perp)$. This behaviour can be also observed in a more general context involving full multislice-based (rather than just first Born approximation based) numerical simulations presented in Figs.2 and 3. Equation (13) constitutes the theoretical basis for the proposed vCTF method for virtual optical sectioning.

## 2.3. Back-propagated vCTF signal in the case of a monomorphous object

Let us now consider the case of the so called "homogeneous" or "monomorphous" samples, in which the attenuation is proportional to the phase shift, i.e. $a_m(\mathbf{r}_\perp) = \sigma \varphi_m(\mathbf{r}_\perp)$, where the dimensionless constant $\sigma$ is independent of $\mathbf{r}_\perp$ and $m$ [32,33]. In the case of X-rays, such relationship holds, for example, for objects consisting predominantly of a single material (with variable density). In the case of electron imaging, such relationship may be associated with the so-called "phenomenological absorption" [34]. The previous case of a pure phase object can be formally obtained here in the limit of $\sigma = 0$. Substituting the relationship $a_m(\mathbf{r}_\perp) = \sigma \varphi_m(\mathbf{r}_\perp)$ into eqs.(5) and using the identities $\sigma \cos\alpha - \sin\alpha = \sqrt{1+\sigma^2}\sin(\omega-\alpha)$, $\sigma\sin\alpha + \cos\alpha = \sqrt{1+\sigma^2}\cos(\omega-\alpha)$, which hold exactly with $\sin\omega = \sigma/\sqrt{1+\sigma^2}$, $\cos\omega = 1/\sqrt{1+\sigma^2}$ and $\omega = \tan^{-1}(\sigma)$, we obtain

$$\begin{cases} \hat{a}(\mathbf{q}_\perp, z) = \sqrt{1+\sigma^2} \sum_{m=1}^{M} \sin[\pi\lambda(z_m - z)q_\perp^2 + \omega]\hat{\varphi}_m(\mathbf{q}_\perp) \\ \hat{\varphi}(\mathbf{q}_\perp, z) = \sqrt{1+\sigma^2} \sum_{m=1}^{M} \cos[\pi\lambda(z_m - z)q_\perp^2 + \omega]\hat{\varphi}_m(\mathbf{q}_\perp) \end{cases}. \quad (14)$$

When $z = z_m$, eqs.(14) imply that $\hat{a}(\mathbf{q}_\perp, z_m) = \sigma\sum_{m=1}^{M}\hat{\varphi}_m(\mathbf{q}_\perp)$ and $\hat{\varphi}(\mathbf{q}_\perp, z_m) = \sum_{m=1}^{M}\hat{\varphi}_m(\mathbf{q}_\perp)$, as expected. In the case of a single scattering centre, we get $\hat{K}(\mathbf{q}_\perp, z) = 2\hat{a}_m(\mathbf{q}_\perp, z) = 2\sqrt{1+\sigma^2}\sin[\pi\lambda(z_m - z)q_\perp^2 + \omega]\hat{\varphi}_m(\mathbf{q}_\perp)$. This equation is analogous to eq.(6) presented above in the pure phase case. The equation for the vCTF reconstructed signal here can be defined as follows:

$$\hat{f}_\sigma(\mathbf{q}_\perp, z) \equiv \frac{\cos\omega}{2\sin[\pi\lambda(z-D)q_\perp^2 + \omega]} \hat{K}(\mathbf{q}_\perp, D). \quad (15)$$

When $z = z_m$, eq.(15) gives $\hat{f}_\sigma(\mathbf{q}_\perp, z_m) = \hat{\varphi}_m(\mathbf{q}_\perp)$. When $z = D$, eq.(15) becomes $\hat{f}_\sigma(\mathbf{q}_\perp, D) = \sigma^{-1}\hat{a}_m(\mathbf{q}_\perp, D)$, i.e., unlike eq.(7), eq.(15) does not have a singularity at the image plane when $\sigma \neq 0$. However, the denominator of eq.(15) can still be equal to zero at some points and, hence, a regularization, e.g. similar to that used in eq.(8), is necessary in general:

$$\hat{f}_{\sigma,\varepsilon}(\mathbf{q}_\perp, z) \equiv \frac{\cos\omega \sin[\pi\lambda(z-D)q_\perp^2 + \omega]}{2\sin^2[\pi\lambda(z-D)q_\perp^2 + \omega] + \varepsilon} \hat{K}(\mathbf{q}_\perp, D), \quad (16)$$

where $0 < \varepsilon \ll 1$ as before. It is easy to see that when $\omega \ll 1$, $\hat{f}_{\sigma,\varepsilon}(\mathbf{q}_\perp, z) \cong \hat{f}_\varepsilon(\mathbf{q}_\perp, z)$, with $\hat{f}_\varepsilon(\mathbf{q}_\perp, z)$ defined by eq.(8). This means that eqs.(15) and (16) smoothly transition into the pure phase eqs.(7) and (8), respectively, when the absorption tends to zero.

Similar to eq.(10), in a vicinity of $z_m$ we have $\hat{f}_{\sigma,\varepsilon}(\mathbf{q}_\perp, z) \cong \hat{\varphi}_m(\mathbf{q}_\perp, z)$, and, correspondingly, in real space

$$f_\sigma(\mathbf{r}_\perp, z) \cong \left( \frac{\sin\{\pi r_\perp^2 / [\lambda(z-z_m)]\}}{\lambda(z-z_m)} + \frac{\sigma\cos\{\pi r_\perp^2 / [\lambda(z-z_m)]\}}{\lambda(z-z_m)} \right) * \varphi_m(\mathbf{r}_\perp). (17)$$

Comparing this with eq.(11), we see that the first term in brackets on the right-hand side of eq.(17) corresponds to $f(\mathbf{r}_\perp, z)$. The second term in the brackets introduces an asymmetry and shift of the maximum with respect to the point $z = z_m$ in the reconstructed signal. A sample profile of the LPSF $f_{\sigma,\varepsilon}(0,0,z)$ obtained using eq.(16) with $\sigma = \varepsilon = 0.1$ is shown in Fig.2 (orange curve). Comparing this result with the corresponding result obtained under the same conditions with eq.(8), i.e. with no absorption, one can see that the inclusion of absorption has indeed led to a slight asymmetry in the shape of the peak near the atomic position and to a small shift of the peak along the optic axis.

Importantly, it has been shown previously that eqs.(14) hold for a broader class of samples that may not satisfy the WOA, provided that the distribution of the refractive index in a sample can be represented as a sum of a small function and a slowly varying function [35,36]. For such samples, the reconstruction based on eq.(15) should still work well.

## 2.4. Reconstruction of the 3D distribution of refractive index

We shall now proceed with the solution of the problem of reconstruction of the 3D distribution of the refractive index using the vCTF method. For the sake of brevity, we will again present the derivation for the pure phase case, but will also give equivalent formulae for monomorphous objects after that. Let us first assume that the phases $\varphi_m(\mathbf{r}_\perp)$ of individual localized scatters in the imaged sample can be reconstructed from the 2D contrast $K(\mathbf{r}_\perp, D)$ of a single image obtained at the defocus distance $z = D$. This can be achieved, as described above after eq.(13), by finding the peaks $f_\varepsilon(\mathbf{r}_\perp, z_m) \cong \varphi_m(\mathbf{r}_\perp)$ of the vCTF solution $f_\varepsilon(\mathbf{r}_\perp, z)$ defined by eq.(8). Since $\varphi_m(\mathbf{r}_\perp) = (2\pi/\lambda)\int \delta_m(\mathbf{r}_\perp - \mathbf{r}_{m,\perp}, z - z_m)dz$, we now have to reconstruct the distribution of the local refractive index from its linear integral. This is possible e.g. when the refractive index of an individual scatterer, $\delta_m(\mathbf{r}_\perp - \mathbf{r}_{m,\perp}, z - z_m)$, is known to be rotationally symmetric, i.e. invariant with respect to rotations around an axis parallel to $\mathbf{r}_\perp$. Let us assume that the axis of symmetry coincides with the coordinate $y$. Then $\delta_m(\mathbf{r}_\perp - \mathbf{r}_{m,\perp}, z - z_m)$ can be found from its linear integral along $z$ using the Abel transform formula [4]:

$$\delta_m(\mathbf{r} - \mathbf{r}_m) = \frac{-\lambda}{2\pi^2 \rho} \frac{\partial}{\partial \rho} \int_\rho^\infty \frac{\rho' \varphi_m(\rho', y)}{(\rho'^2 - \rho^2)^{1/2}} d\rho',\ \rho = \sqrt{x^2 + (z - z_m)^2}. \quad (18)$$

Furthermore, if $\delta_m(\mathbf{r} - \mathbf{r}_m)$ is 3D spherically symmetric and sufficiently slowly varying, the above equation can be approximated by a simpler one [14]:

$$\delta_m(\mathbf{r} - \mathbf{r}_m) \cong [\lambda/(2\pi w_m)]\varphi_m(\mathbf{r}_\perp), \quad (19)$$

where $w_m$ is the "width" of the distribution $\delta_m(\mathbf{r} - \mathbf{r}_m)$. For example, when $\delta_m(\mathbf{r}) = \exp[-|\mathbf{r}|^2/(2\sigma_m^2)]$, eq.(19) gives exactly the same result as eq.(18) when $w_m = (2\pi)^{1/2}\sigma_m$.

When a full distribution of the refractive index is described by eq.(1), the reconstruction of each individual scatterer can proceed independently using eq.(19), with the whole sample reconstruction being a superposition of the results for the individual scatterers:

$$\delta(\mathbf{r}) \cong [\lambda/(2\pi w)] f_\varepsilon(\mathbf{r}_\perp, z), \quad (20)$$

where we assumed for simplicity that the width of all individual scatterers is approximately the same, $w_m \cong w$. It is explained in [15] that the violation of the latter condition usually leads to benign reconstruction artefacts and, in particular, it does not affect the 3D localization of individual scatterers. According to eq.(20), at the location, $\mathbf{r} = \mathbf{r}_m$, of any individual scatterer, we will have $\delta(\mathbf{r}_m) \cong \delta_m(0) \cong [\lambda/(2\pi w_m)]\varphi_m(\mathbf{r}_{\perp,m}) \cong [\lambda/(2\pi w)] f_\varepsilon(\mathbf{r}_{\perp,m}, z_m)$. Equation (20), together with eq.(8), provides a basis for the vCTF VOS method from a single defocused image, $I(\mathbf{r}_\perp, D)$. This is achieved by back-propagating the image contrast $K(\mathbf{r}_\perp, D) = 1 - I(\mathbf{r}_\perp, D)/I_{in}$ to different $z$ according to eq.(8) and finding the maxima (peaks) of the function $f_\varepsilon(\mathbf{r}_\perp, z)$. According to eq.(20), the latter result will correspond to the peaks of the refractive index in the sample.

The vCTF method can also be obtained for VOS of monomorphous objects. In this case, the function $f_{\sigma,\varepsilon}(\mathbf{r}_\perp, z)$ defined in eq.(16) should be used in the right-hand side of eq.(20) instead of $f_\varepsilon(\mathbf{r}_\perp, z)$. Once the real increment of the refractive index is obtained using this modified equation, the imaginary part can be found as $a_m(\mathbf{r}_\perp) = \sigma\varphi_m(\mathbf{r}_\perp)$, where the proportionality coefficient $\sigma$ is supposed to be known *a priori*.

For some samples and illumination directions, the "virtual optical sectioning" in accordance with eq.(20) may not be possible because, for example, for a given illumination direction, two or more individual scattering centres are located on the same line parallel to the optical axis, i.e. $\mathbf{r}_{m_1,\perp} = \mathbf{r}_{m_2,\perp}$ for $m_1 \neq m_2$, while $z_{m_1} \neq z_{m_2}$. In such cases one may have to resort to a tomographic-style reconstruction from multiple defocus images collected at different illumination directions (or, equivalently, at a fixed illumination direction, but different 3D rotational positions of the sample). Similarly to the derivation presented in [14,15], it is relatively straightforward to show that in the latter case the reconstruction formula eq.(20) can be simply averaged over all available illumination directions to give the desired result. The relevant tomographic reconstruction formula is

$$\delta(\mathbf{r}) \cong \frac{\lambda}{8\pi^2 w} \int_0^{2\pi}\int_0^{\pi} f_\varepsilon(\mathbf{r}_{\theta,\varphi,\perp}, z_{\theta,\varphi}) |\sin\varphi| \, d\theta d\varphi, \qquad (21)$$

where the function $f_\varepsilon(\mathbf{r}_{\theta,\varphi,\perp}, z_{\theta,\varphi})$ corresponds to the inverse Fourier transform of eq.(8) in the rotated coordinates $\mathbf{r}_{\theta,\varphi} \equiv (x_{\theta,\varphi}, y_{\theta,\varphi}, z_{\theta,\varphi}) = (\mathbf{r}_{\theta,\varphi,\perp}, z_{\theta,\varphi})$. A choice of 3D Euler rotation angles in eq.(21) is unimportant. In its present form, eq.(21) is written with respect to the initial rotation by the angle $\theta$ around the $y$ axis, followed by a rotation by the angle $\varphi$ around the new $x_\theta$ axis:

$$\begin{cases} x_{\theta,\varphi} = x_\theta = x\sin\theta + z\cos\theta \\ y_{\theta,\varphi} = y_\theta \cos\varphi + z_\theta \sin\varphi = y\cos\varphi - x\cos\theta\sin\varphi + z\sin\theta\sin\varphi \\ z_{\theta,\varphi} = -y_\theta \sin\varphi + z_\theta \cos\varphi = -y\sin\varphi - x\cos\theta\cos\varphi + z\sin\theta\cos\varphi. \end{cases} \qquad (22)$$

Note that the factor $|\sin\varphi|$ inside the integral in eq.(21), together with the factor $4\pi$ in the denominator, correspond to the case with the available illumination directions being uniformly distributed over $\theta \in [0,\pi)$ and $\varphi \in [0, 2\pi)$. In principle, any set of illumination directions and/or rotational positions of the sample can be used for the 3D reconstruction of the refractive index with the help of eq.(21), provided that it is suitably modified for a particular sampling scheme. For example, in an imaging experiment where a discrete set of $n_a$ illumination directions are uniformly distributed over the unit sphere in 3D, the factor $|\sin\varphi|/(4\pi)$ in eq.(21) should be replaced by $1/n_a$. It is, however, important to ensure that the input set of illumination directions is sufficiently "rich" to suppress the contribution of multiple scattering. A more detailed discussion of this problem can be found in a similar context in [15].

As in the case of eq.(20) above, eq.(21) can also be easily modified for 3D reconstruction of the complex refractive index in monomorphous objects by replacing the function $f_\varepsilon(\mathbf{r}_{\theta,\varphi,\perp}, z_{\theta,\varphi})$ in the right-hand side of eq.(21) with the function $f_{\sigma,\varepsilon}(\mathbf{r}_{\theta,\varphi,\perp}, z_{\theta,\varphi})$ defined by eq.(16) at each illumination direction.

## 3. Numerical example

In this section, we present a numerical example of vCTF reconstruction of a platinum-iron nanoparticle [37] containing 5,107 Pt atoms and 5,356 Fe atoms, placed on a 100 Å thick amorphous carbon substrate containing 90,253 C atoms. The numerical simulations were performed using open-source software [17]. For the simulations, the whole structure containing the nanoparticle and the substrate was placed into a virtual cubic volume $Q_{200}$ with 200 Å sides, which was located in the positive octant of the Cartesian coordinates in 3D, with the sides parallel to the coordinate axes and one corner located at the origin of coordinates. A 3D rendering of the nanoparticle and the substrate in one 3D orientation is shown in Fig.5(a). The 3D image was obtained using the Vesta software [38] with the input text file (in the XYZ format) containing the 3D positions of all atoms. The structure was illuminated by a plane monochromatic electron beam with E = 200 keV ($\lambda \cong 0.025$ Å) propagating along the $z$ axis. The particle with the substrate was subsequently rotated in 3D to 360 different pseudo-random orientations with the corresponding directional vectors uniformly distributed on the unit sphere. The effective centre of rotation was at $(x, y, z) = (100\text{Å}, 100\text{Å}, 100\text{Å})$.

At each orientation, the transmission of the electron wave through the volume $Q_{200}$ containing the particle and the substrate was calculated using a multislice-based algorithm [18-20]. In these calculations, an effective objective aperture of 40 mrad was assumed, achievable in aberration-corrected TEM. The effect of thermal motion of atoms was included in the simulations via a Debye-Waller factor with the root-mean square displacement of 0.085 Å at 300 K. The complex amplitude of the transmitted electron wave was then propagated in free space from the exit plane $z = 200$ Å to the defocus plane. For different orientations of the imaged structure, the defocus planes were located at different $z$ positions which were uniformly randomly distributed between $z = 300$ Å and $z = 350$ Å. For each defocused image with $1{,}024 \times 1{,}024$ pixels, pseudo-random Poisson shot noise with a mean corresponding to 59 electrons per Å$^2$ (approximately 2.25 electrons per pixel) was also simulated. A typical 2D defocused image at one of the 3D orientations of the particle is shown in Fig.5(b).

We then performed a vCTF reconstruction from the 360 simulated noisy defocused images using eq.(8) and eq.(21) with a Tikhonov regularization parameter $\varepsilon = 0.1$. A sample 2D cross-section, corresponding to the $(x, y)$ plane at $z = 100$ Å, through the reconstructed 3D distribution of the electrostatic potential $V(\mathbf{r}) = 2E\delta(\mathbf{r})$ is shown in Figs.5(c) and (d). It is easy to see the peaks corresponding to many individual atoms in these figures. By splitting the part of the reconstruction volume $Q_{200}$ containing the nanoparticle into non-overlapping cubes with a side of 1.7 Å, finding the positions of the maxima of the reconstructed potential inside each such small cube and selecting only the maxima higher than 10.7 V, we found 10,464 candidate atomic locations. A simple pair-wise comparison of these reconstructed locations with the 10,463 actual atomic positions in the Fe-Pt nanoparticle resulted in 10,373 matches with the average distance of 0.12 Å and the maximum distance of 0.88 Å between the original and the reconstructed atomic locations. Thus, 99.14 % of atoms in the particle were located with sub-Å precision in this reconstruction. The result contained 90 false negatives (i.e. missed atoms) (0.86 %) and 91 false positives (i.e. peaks of the reconstructed potential not associated with any actual atoms in the particle) (0.87 %). All Pt atoms were correctly located, while 90 Fe atoms were missed in this reconstruction. Of the first 5,107 highest identified peaks in the reconstructed electrostatic potential, 5,073 corresponded to locations of Pt atoms in the original XYZ file of the nanoparticle, and only 34 of such peaks corresponded to the original locations of Fe atoms. This means that the vCTF method has demonstrated very high specificity (99.3% of correct identifications) with respect to the two atom types in the imaged nanoparticle.

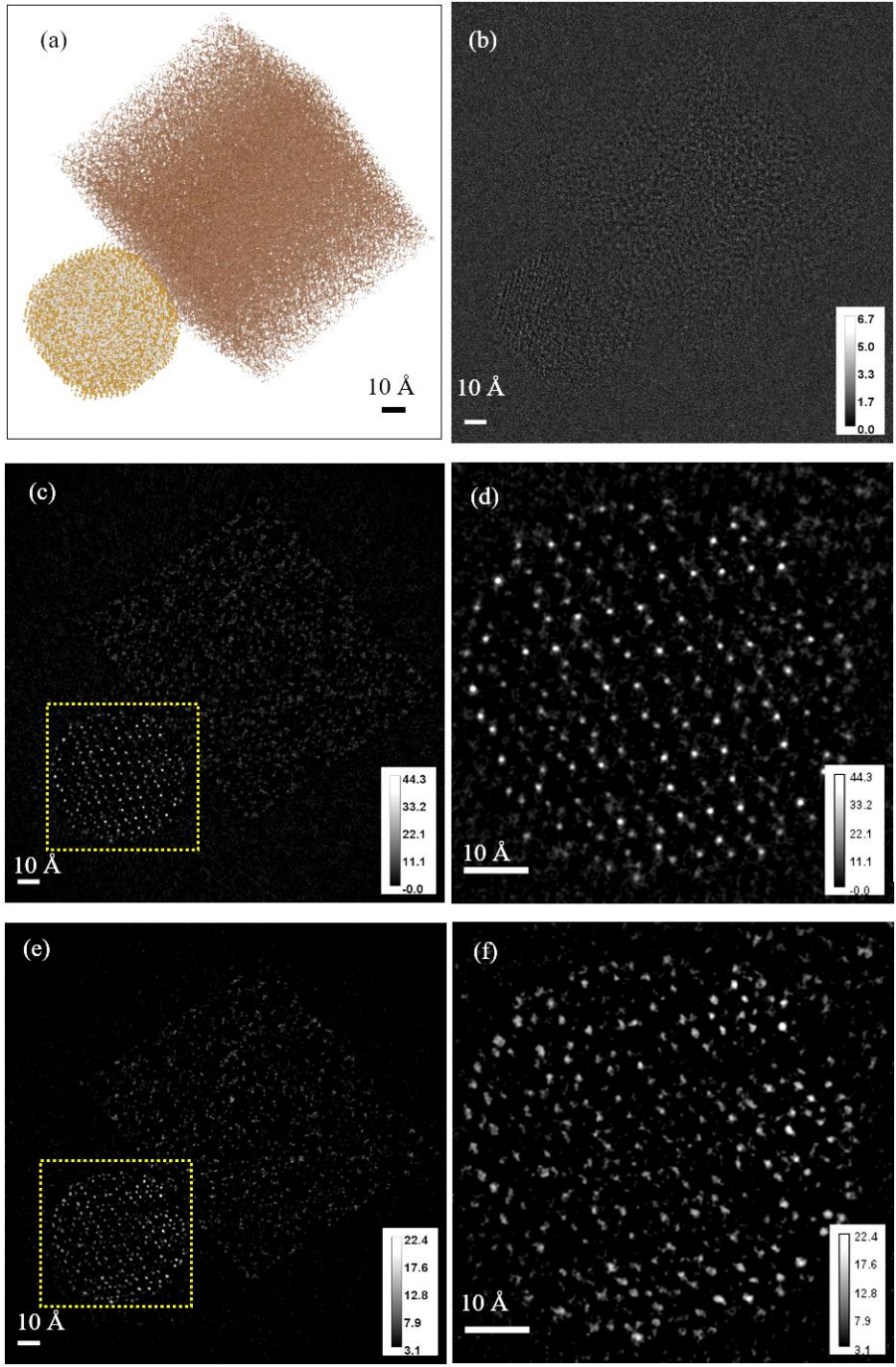

Fig. 5. (a) 3D rendering of the Fe-Pt nanoparticle on a C substrate used in the simulations included in section 3. (b) A typical simulated defocused image with Poisson noise. (c) 2D cross-section at the level $z = 100$ Å through the 3D distribution of the electrostatic potential $V(\mathbf{r}) = 2E\delta(\mathbf{r})$ reconstructed using eq.(8) and eq.(21) from 360 defocused images with Poisson noise corresponding to a mean of 2.25 electrons per pixel. (d) Zoomed (×3) version of the area shown by yellow rectangle in (c). (e) 2D cross-section at $z = 100$ Å through the potential reconstructed using the cCTF method from the same 360 defocused images. (f) Zoomed (×3) version of the area shown by yellow rectangle in (e). The grey-scale levels in (c)-(d) are in volts.

We have also performed the cCTF reconstruction from the same 360 noisy defocused images of the Fe-Pt nanoparticle. According to common practice, we used here the conventional CTF correction with a fixed distance corresponding to the distance between the centre of the imaged structure and the image plane at each illumination direction. Intrinsic to this method, the DOF is not taken into account. In other words, the positions of atoms along the illumination direction inside the molecule are ignored in each back-projection from a single defocused image. In the present case with $\lambda \cong 0.025 \text{Å}$ and the aperture of 40 mrad, we had the spatial resolution $\Delta \cong 1/q_{\perp,\max} \cong 0.025\text{Å}/\tan(0.04) \cong 0.625\text{Å}$ and hence the $\text{DOF} = \Delta^2/(2\lambda) \cong 7.8\text{Å}$. This DOF is substantially smaller than the size of the nanoparticle (70 Å) and, therefore, as explained in the Introduction, one could expect noticeable blurring of the cCTF-reconstructed potential in the vicinity of atoms located at distances exceeding the DOF from the centre of the particle. The 2D cross-section at $z = 100$ Å through the cCTF-reconstructed 3D potential is shown in Figs.5(e) and (f). It is easy to see in Fig.5(f), in particular, that the peaks of the reconstructed potential near the atoms located away from the centre of rotation are indeed blurry and are lower compared to the vCTF-reconstructed peaks at the same locations in Fig.5(d). As a result of this lower resolution and lower SNR in the cCTF-reconstructed 3D distribution of the electrostatic potential, compared to the previous vCTF reconstruction, the same peak-finding procedure as above resulted in the successful localization of only 7,510 out of 10,463 atoms in the nanoparticle. This corresponded to 71.78 % of successful localizations, compared to over 99 % in the case of vCTF reconstruction above. This difference in the quality of the reconstruction between the cCTF and vCTF methods confirms that taking into account the in-molecule propagation (i.e. the finite DOF or the curvature of the Ewald sphere) does improve the spatial resolution and the SNR in the 3D reconstruction of the refractive index from defocused images obtained with a shallow DOF.

As the DOF in this case was substantially larger than the average distance between atomic planes in the nanoparticle, and the nanoparticle contained a relatively large number of scattering centres conducive to multiple scattering, the VOS vCTF reconstruction from a single defocused image proved to be unsuccessful. Our attempted vCTF reconstructions from a single image did show longitudinal variations of the reconstructed potential, i.e. different $z$-sections did look different, as would be expected in a VOS. However, it proved difficult to establish a correspondence between the atomic peaks in the VOS reconstruction with the true positions of the atoms in the corresponding planes orthogonal to the illumination direction. The fact that we were previously able to successfully perform the VOS of the aspartate molecule imaged with a comparable DOF shows that in order for such a procedure to work in practice, the sample needs to be sufficiently sparse so that multiple scattering affects, together with a DOF exceeding the inter-atomic distances, would not spoil the reconstruction. The conditions sufficient for a successful application of vCTF VOS may need to be investigated further in the future, preferably in the context of a particular class of imaging problems.

As explained in the Introduction, in the presence of strong multiple scattering, it may be necessary to collect many transmission images at different orientations of the sample in order to obtain an accurate 3D reconstruction of the internal structure, as was achieved using 360 different views of the Fe-Pt nanoparticle above. Note that the latter number of different views is still lower than the approximately 503 views required according to the Nyquist sampling conditions in CT in order to achieve a 3D reconstruction with an approximately isotropic spatial resolution of ~0.625 Å (as above). Despite using a smaller number of images than required by the conventional CT sampling conditions, the above vCTF reconstruction was able to locate over 99 % of atoms in the nanoparticle with the average positional accuracy of approximately 0.1 Å.

## 4. Conclusions

In the present paper, we have developed and tested the vCTF method capable of virtual optical sectioning and 3D tomographic reconstruction of internal structure of objects from planar transmission images. The method relies on a sufficiently shallow DOF as a means for achieving non-trivial spatial resolution along the illumination direction in the reconstruction. While this spatial resolution does depend on the extent of the DOF, a "super-resolution" can sometimes be achieved by localizing the peaks of the reconstructed signal along the back-propagation direction. The largest obstacle to the application of this method in the VOS form, i.e. to the 3D imaging of a sample from a single defocused image, is represented by multiple scattering in the sample. In the simplest case, if two or more such centres are located along the same illuminating ray, they can potentially "shade" each other in the defocused image. In order to achieve unambiguous 3D reconstruction in such cases, it may be necessary to collect several images at different illumination directions or several rotational positions of the sample. Such an imaging scheme is similar to that used in conventional CT methods. However, we have shown that the shallow depth of field can still provide benefits in this case by increasing the SNR and removing the blurring in the reconstructed refractive index on the periphery of the sample, i.e. in the regions of the sample that are located more than one DOF away from the centre of CT rotation. The tomographic variant of the vCTF method has also been explicitly developed above, with a simple reconstruction formula presented that is based on angular averaging of all partial reconstructions (back-propagation traces) obtained from individual defocused images. The proposed method may find application in transmission imaging with electrons, soft X-rays, visible light, ultrasound and any other radiation or matter waves, provided that the DOF is shallower than the thickness of the sample along the illumination direction. The vCTF method has been implemented in open-source software that is publicly available for download from GitHub [17].

**Acknowledgments.** The authors would like to thank J. Barthel, H. G. Brown, S. D. Findlay and D. M. Paganin for helpful discussions related to the present work.

**Disclosures.** The authors declare no conflicts of interest.

**Data availability.** Data underlying the results presented in this paper are not publicly available at this time but may be obtained from the authors upon reasonable request.